# On the characteristic structure of the adjoint Euler equations and the analytic adjoint solution of supersonic inviscid flows

**Carlos Lozano [1,*] and Jorge Ponsin [2]**

[1] Computational Aerodynamics, National Institute of Aerospace Technology (INTA), Spain; lozanorc@inta.es
[2] Computational Aerodynamics, National Institute of Aerospace Technology (INTA), Spain; ponsinj@inta.es

**Abstract:** The characteristic structure of the two-dimensional adjoint Euler equations is examined. The behavior is similar to that of the original Euler equations, but with the information travelling in the opposite direction. The compatibility conditions obeyed by the adjoint variables along characteristic lines are derived. It is also shown that adjoint variables can have discontinuities across characteristics and the corresponding jump conditions are obtained. It is shown how this information can be used to obtain exact predictions for the adjoint variables, particularly for supersonic flows. The approach is illustrated by the analysis of supersonic flow past a double wedge airfoil, for which an analytic adjoint solution is obtained in the near-wall region. The solution is zero downstream of the airfoil and piecewise constant around it except across the expansion fan, where the adjoint variables change smoothly while remaining constant along each Mach wave within the fan.

---

## List of symbols

|  |  |
|---|---|
| $A$, $B$ | coefficient matrices of a system of PDE |
| $c$ | Speed of sound |
| $F_x$, $F_y$ | Inviscid flux Jacobian matrices |
| $I_1, I_2, I_3, I_4$ | Linearized cost functions due to point source perturbations |
| $K_H, K_S, K_\pm$ | Flow compatibility conditions along characteristic lines |
| $l_\lambda$ | Left eigenvector of $B-\lambda A$ |
| $M$ | Local Mach number |
| $\hat{n}_\lambda$ | unit normal vector to characteristic line with slope $\lambda$ |
| $q$ | Modulus of velocity vector |
| $r_\lambda$ | Right eigenvector of $B-\lambda A$ |
| $R_\lambda$, | Riemann invariant of eigenvalue $\lambda$ |
| $R_1^\psi, R_2^\psi, R_\pm^\psi$ | Adjoint Riemann invariants |
| $(u,v)$ | Cartesian components of the fluid velocity vector |
| $\alpha$ | $\sqrt{M^2-1}$ |
| $\gamma$ | Ratio of specific heats (adiabatic index) |
| $\delta$ | Tangent derivative along a characteristic line |
| $\theta$ | Local streamline direction |
| $\kappa$ | Normalization constant of the drag coefficient |
| $\lambda$ | Eigenvalue of characteristic equation, slope of characteristic line |
| $\mu$ | Mach angle |
| $\nu$ | Prandtl-Meyer function |
| $\tau$ | Trailing edge half-angle of symmetrical double-wedge airfoil |
| $\Psi$ | Adjoint state |
| $\psi_1, \psi_2, \ldots$ | Components of the adjoint state |

## 1. Introduction

The adjoint equations were introduced for design optimization in the field of computational aerodynamics by Jameson [1, 2], and have been since extended to a variety of applications such as optimal shape design [3] of aircrafts [4], ships [5] and automobiles [6], error estimation and goal-oriented mesh adaptation [7] and stability analysis [8], among many others.

Of particular interest is the application of adjoint methods to the shape design of supersonic aircraft. Early applications include the aerodynamic optimization of a supersonic transport configuration [9] and the CAD-based shape optimization of a reentry capsule in hypersonic flow [10]. More recently, and in an attempt to address environmental concerns of commercial supersonic flight, a program was initiated to apply adjoint methods to the reduction of the sonic boom. The adjoint method was used in [11] to investigate the influence of geometry modifications on the pressure distribution at remote locations, while [12, 13, 14] apply adjoint methods to compute the sensitivities of functionals based on the equivalent area distribution.

In all of these applications, the adjoint method is used to compute the linear sensitivity of a cost or objective function with respect to a number of independent variables defining perturbations of the flow. These can be design variables in shape optimization applications or numerical errors or tolerances in mesh adaptation or uncertainty quantification applications. From a mathematical perspective, the adjoint variables appear as Lagrange multipliers enforcing the flow equations in a variational analysis of the cost function. The vanishing of the variation requires the multipliers to obey a partial differential equation (the adjoint equation) and appropriate boundary conditions. Aside from their mathematical meaning, the adjoint variables also carry physical significance as giving the influence of point-sources (Green's functions) of mass, momentum and energy, respectively, on the objective function [15].

An important step in the application of adjoint methods is the development and verification of adjoint codes. In recent times, a program has been initiated by several authors to produce analytic predictions for the adjoint equations that can be used for those purposes (but also to gain understanding on the behavior of the adjoint equations). The connection of the adjoint variables to Green's functions has made it possible to generate exact adjoint solutions for quasi-1D inviscid flows [16, 17] and two-dimensional (2D) incompressible inviscid flows [18, 19], and outline the solution for 2D compressible subcritical inviscid flows [20]. For inviscid 2D/3D flows, entropy variables offer another exact solution for quantifying the net entropy flux across boundaries [21, 22]. A closely related solution, corresponding to the near-field computation of aerodynamic drag, was recently discovered by the authors [23]. For viscous flows, exact adjoint solutions for the Navier-Stokes equations have been proposed for both laminar [24] and turbulent [25] boundary layers. Lastly, two recent papers [26, 27] derived ordinary differential equations that must be obeyed by the adjoint solutions along characteristic lines which can be used as a verification tool for numerical adjoint fields.

This paper considers the characteristic structure of the adjoint Euler equations with particular emphasis in supersonic flow. For simplicity, the analysis is restricted to two dimensional flows. Steady inviscid compressible flow in two dimensions obeys the Euler equations

$$R(U)=\nabla\cdot\vec{F}(U)=\partial_x F_x+\partial_y F_y=0 \tag{1}$$

where

$$U=\begin{pmatrix}\rho\\ \rho u\\ \rho v\\ \rho E\end{pmatrix},\quad F_x=\begin{pmatrix}\rho u\\ \rho u^2+p\\ \rho uv\\ \rho uH\end{pmatrix},\quad F_y=\begin{pmatrix}\rho v\\ \rho vu\\ \rho v^2+p\\ \rho vH\end{pmatrix} \tag{2}$$

In eq. (2), $(u,v)$ are the Cartesian components of the velocity, $\rho$ is the density, $p$ is the pressure and $E$ and $H$ are the total energy and enthalpy, respectively. For a perfect gas,

$$p=(\gamma-1)\rho\left(E-\frac{1}{2}q^2\right) \tag{3}$$

and

$$\rho H=\rho E+p \tag{4}$$

where $q=\sqrt{u^2+v^2}$ is the modulus of the velocity and $\gamma$ is the ratio of specific heats. The flow equations (1) can be written in quasi-linear form

$$R(U)=(\partial_U F_x,\partial_U F_y)\cdot(\partial_x U,\partial_y U)=0 \tag{5}$$

where $(\partial_U F_x,\partial_U F_y)$ are the inviscid flux Jacobian matrices. Eq. (5) is the starting point for the characteristic analysis of the Euler equations as will be shown momentarily.

The derivation of the adjoint Euler equations is quite standard [28] and will not be repeated here. For definiteness, the focus will be set on an external aerodynamics problem consisting on an airfoil with profile $S$ immersed in a flow with incidence $\theta_\infty$. For this case, it is desired to compute the sensitivities of a cost function measuring aerodynamic drag given as the integral of the pressure along a wall boundary $S$

$$\frac{1}{\kappa}\int_S p\hat{n}\cdot\vec{d}\,ds \tag{6}$$

where $\kappa = \rho_\infty \left|\vec{v}_\infty\right|^2 \ell / 2$ is a normalization constant, $\ell$ is a reference length – the chord length of the airfoil, $\hat{n} = (n_x, n_y)$ is the wall unit normal vector and $\vec{d} = (\cos\theta_\infty, \sin\theta_\infty)$ lies along the inflow direction. The adjoint equations are

$$-(\partial_U F_x)^T \partial_x \Psi - (\partial_U F_y)^T \partial_y \Psi = 0 \tag{7}$$

where $\Psi = (\psi_1, \psi_2, \psi_3, \psi_4)^T$ are the adjoint variables. The adjoint boundary conditions are chosen to eliminate the integral of $\Psi^T (\partial_U F_x, \partial_U F_y) \cdot (n_x, n_y) \delta U$ on the boundaries. This procedure results in dual characteristic boundary conditions (obtained from a locally one-dimensional characteristic decomposition) in the far-field, while at the wall the adjoint variables obey

$$n_x \psi_2 + n_y \psi_3 = (n_x \cos\theta_\infty + n_y \sin\theta_\infty) / \kappa \tag{8}$$

The characteristic structure of eq. (5) is well known and depends on the local Mach number $M$ [29]. For subsonic flow, there is only one family of characteristic lines (the streamlines), while for supersonic flows there are two additional families of characteristics, the Mach lines that are inclined at an angle $\pm\sin^{-1}(1/M)$ to the local flow direction. Along characteristics, the partial differential equations (PDE) that describe the flow reduce to ordinary differential equations (ODE) called compatibility equations. In certain circumstances, notably in the case of two-dimensional irrotational supersonic flow, the compatibility conditions can be integrated and are further reduced to algebraic equations that hold only along the characteristic lines. The information carried by the compatibility conditions can be used to simplify the computation of supersonic flows, one notable practical application being the design of supersonic nozzles [29].

The adjoint equations (7) share the same characteristic structure as the flow equations but with the sign of each characteristic velocity reversed so that the characteristic information travels in the opposite direction [1, 15]. As in the previous case, the adjoint variables obey compatibility conditions along the characteristics (see e.g. [30] and references therein).

The adjoint equations are linear and, for supersonic flows, hyperbolic. As a result of this, the behavior of the analytic supersonic adjoint solutions is strongly constrained by the characteristic structure of the equations. On the one hand, and barring special cases where cost functions are defined along certain lines, not necessarily characteristic, such as in [11], the adjoint variables can only have discontinuities along characteristic lines. The jumps in the adjoint variables along these lines are constrained by jump conditions as will be shown in section 2. Additionally, since in supersonic flow information travels along characteristic lines, supersonic adjoint solutions tend to follow characteristic lines emanating from significant features of the flow or the geometry such as leading and trailing edges [26, 31] , shock feet [32], expansion fan centers [33], nozzle lips [27] etc. (see also section 3), and these trends are clearly visible in adjoint-adapted meshes [34]. This behavior is significant as it is precisely in those regions, where the amplitude of the adjoint solution is usually highest, that the cost function is most sensitive to perturbations.

In this paper, the characteristic structure of the adjoint Euler equations is reviewed and the corresponding compatibility and jump conditions along and across characteristic lines are obtained. Most of what will be said is well-known, but the aim is to present the material in a comprehensive and unified way. It is also shown how this information can be used to obtain quantitative predictions for the adjoint variables in supersonic flow. In this way the results presented in this paper go beyond [26, 27], where the compatibility conditions were derived and checked on several high-quality, numerical adjoint solutions.

The paper is organized as follows. Section 2 is devoted to the review of the derivation of the characteristic structure of the flow and adjoint Euler equations in 2D, including the eigenvalues, characteristic directions and compatibility and jump conditions associated to the characteristic lines. These conditions are applied in section 3 to the analysis of supersonic flow past a diamond airfoil, for which a closed-form analytic adjoint solution, valid in the near wall region, is presented and checked against a numerical adjoint solution on a very fine grid. In section 4, we summarize our findings and make some concluding remarks.

## 2. Characteristic Structure of the Adjoint Euler Equations

This section contains a brief analysis of the characteristic structure of the flow and adjoint Euler equations. The reader is also referred to the recent works [26, 27], where the characteristic equations are obtained from a somewhat different perspective.

### *2.1. Compatibility Conditions: Reduction of the PDE to an ODE*

Both the 2D steady flow and adjoint Euler equations (5) and (7) are written in quasilinear form,

$$A\partial_x u + B\partial_y u = 0 \tag{9}$$

For a general quasilinear system (9), its characteristic structure is determined by the solution of the eigenvalue problem [35]

$$\det(B - \lambda A) = 0 \tag{10}$$

Notice that, since $\det(B-\lambda A) = \det(B^T - \lambda A^T)$, it is clear that the characteristic structure of the flow and adjoint equations is identical.

If $\lambda$ is a real solution of eq. (10), then the planar curve with local slope

$$\frac{dy}{dx} = \lambda \tag{11}$$

is a characteristic curve of the above quasilinear system. Across characteristics, both the solution and its first derivatives may be discontinuous. Furthermore, the original system of equations can be reduced to an ODE along characteristics. Such ODES, one per characteristic curve, are called compatibility conditions, and can be obtained as follows.

Suppose $\lambda$ is a real eigenvalue of $\det(B-\lambda A) = 0$. The matrix $B - \lambda A$ is singular, so there exists a left eigenvector $l_\lambda$ such that

$$l_\lambda^T (B - \lambda A) = 0 \tag{12}$$

Multiplying eq. (9) on the left by $l_\lambda$ and using the properties of the left eigenvector yields a scalar equation

$$l_\lambda^T A\partial_x u + l_\lambda^T B\partial_y u = l_\lambda^T A\partial_x u + \lambda l_\lambda^T A\partial_y u = l_\lambda^T A(\partial_x u + \lambda\partial_y u) = 0 \tag{13}$$

In eq. (13),

$$\partial_x u + \lambda\partial_y u \tag{14}$$

is (proportional to) the (tangent) derivative along the characteristic curve $(x, y(x))$ with local slope $\lambda = dy/dx$, so eq. (13) reduces to the following scalar ODE:

$$l_\lambda^T A\frac{du}{dx} = 0 \tag{15}$$

Eq. (15) is the compatibility condition obeyed along the characteristic line by the solution to the PDE. There is one such compatibility condition along each of the families of characteristics. Besides, if the compatibility condition is integrable, it can be rearranged into the generic form

$$\frac{dR_\lambda}{dx} = 0 \tag{16}$$

where $R_\lambda$, which is conserved along the characteristics $\lambda = dy/dx$, is called a Riemann invariant.

All the above translates immediately to the Euler equations. The adjoint equations, however, are transposed, so a bit of additional work is required. To obtain the adjoint compatibility conditions, the eigenvalue problem $\det(B - \lambda A) = 0$ is considered, where $(A, B) = (\partial_U F_x, \partial_U F_y)$ . For real $\lambda$ one can define, in addition to the left eigenvector, a right eigenvector $r_\lambda$ such that

$$(B - \lambda A)r_\lambda = 0 \tag{17}$$

Multiplying the adjoint equation on the left by $r_\lambda$ and using $Br_\lambda = \lambda Ar_\lambda$ yields the adjoint compatibility condition

$$r_\lambda^T(A^T\partial_x\Psi + B^T\partial_y\Psi) = r_\lambda^T A^T(\partial_x\Psi + \lambda\partial_y\Psi) = r_\lambda^T A^T\frac{d\Psi}{dx} = 0 \tag{18}$$

Under generic circumstances, it is unlikely that the adjoint compatibility conditions can be integrated. However, if the coefficients $r_\lambda^T A^T$ are constant along a certain characteristic line or in regions of constant flow, the quantities $r_\lambda^T A^T\Psi$ (that we will call Riemann functions in what follows) are Riemann invariants for the adjoint equations along $\lambda$ characteristics.

### 2.2. *Discontinuities Across Characteristics: Jump Conditions*

Across characteristics, both the solution and its first derivatives may be discontinuous [35], but the structure of the equations impose matching or jump conditions on the solution and its derivatives. Let us consider a characteristic line associated to the real eigenvalue $\lambda$. Assuming that the Jacobian matrices ($A$,$B$) are continuous (the case of discontinuous Jacobians corresponds to shocks or slip lines that have a singular treatment of their own [17, 36, 37] and will not be further considered here) and distinguish two cases depending on whether $\Psi$ is continuous or not.

If $\Psi$ is continuous, the tangent derivatives across the characteristic line are also continuous and, by decomposing the equation in a local tangent and normal frame relative to the characteristic and taking differences across the characteristic line yields

$$\hat{n}_\lambda \cdot (A,B)^T [\![ \partial_n \Psi ]\!]_\lambda = 0 \tag{19}$$

where $\hat{n}_\lambda \sim (-\lambda, 1)$ is the normal vector to the characteristic line.

On the other hand, if $\Psi$ is discontinuous across the characteristic line, its behavior across the discontinuity is given by the following jump conditions [35]

$$\hat{n}_\lambda \cdot (A,B)^T [\![ \Psi ]\!]_\lambda = 0 \tag{20}$$

The proof of eq. (20) is identical to that of the Rankine-Hugoniot conditions for shocks, involving the integration of the adjoint equation against a test function on a suitably chosen domain divided in two by the line of discontinuity.

Eq. (20) is equivalent to $(B - \lambda A)^T [\![ \Psi ]\!]_\lambda = 0$, so the jump in the adjoint variables across a characteristic line must be a left eigenvector of $B - \lambda A$.

Notice that, in either eqs. (19) and (20), the coefficient matrix of the linear system $\hat{n}_\lambda \cdot (A,B)^T \sim (-\lambda A + B)^T$ is not invertible along the characteristic, so the above equations admit non-trivial solutions. Conversely, if the line is not characteristic, then neither eqs. (19) nor (20) admit non-trivial solutions.

*2.3. Relation of Jumps Across Characteristic Lines and Compatibility Conditions*

The jump conditions give a series of linear combinations of adjoint variables that are continuous across the characteristic line. Of course, not all of these are linearly independent (otherwise, the adjoint variables would be continuous across the characteristic). In fact, $\hat{n}_\lambda \cdot (A,B)^T$ has rank $N - k$, where $N$ is the number of equations and $k$ is the multiplicity of the eigenvalue. This can be understood with the following result, which also sheds light onto the jump conditions. Pick a characteristic line with eigenvalue $\lambda_i$ and multiplicity $k_i$. The corresponding jump conditions are

$$(\lambda_i A^T - B^T) [\![ \Psi ]\!]_{\lambda_i} = 0 \tag{21}$$

Now multiply (21) on the left by the right eigenvector corresponding to one of the $N - k_i$ remaining eigenvalues,

$$r_{\lambda_j}^T (\lambda_i A^T - B^T) [\![ \Psi ]\!]_{\lambda_i} = 0 = (\lambda_i - \lambda_j) r_{\lambda_j}^T A^T [\![ \Psi ]\!]_{\lambda_i} \tag{22}$$

Now, since $\lambda_i \neq \lambda_j$, the jump conditions can be written as

$$r_{\lambda_j}^T A^T \left[\!\left[\Psi\right]\!\right]_{\lambda_i} = 0 \tag{23}$$

for $j \neq i$. Now recall that along a characteristic $j$, the adjoint variables obey the compatibility condition $r_{\lambda_j}^T A^T \delta\Psi = 0$, from where the Riemann functions $r_{\lambda_j}^T A^T \Psi$ can be defined (which turn out to be Riemann invariants along the $j$ characteristic under particular circumstances). It can thus be seen that the jump conditions across one characteristic line are equivalent to the statement that the Riemann functions associated to characteristic lines that cross the given characteristic are continuous.

*2.4. Characteristic Lines and Compatibility Conditions for the Adjoint Euler Equations*

For the 2D Euler equations, the characteristic equation (10) has the following solutions [29]:

$$\lambda_0 = \frac{v}{u} \tag{24}$$

(double) and

$$\lambda_\pm = \frac{uv \pm c^2\alpha}{u^2 - c^2} = \frac{v\alpha \pm u}{u\alpha \mp v} \tag{25}$$

where $c = \sqrt{\gamma p / \rho}$ is the speed of sound, $\alpha = \sqrt{M^2 - 1}$ and $M = q / c$ is the local Mach number. The characteristic lines corresponding to $\lambda_0$ are the flow streamlines. The associated compatibility conditions (15) can be arranged in the following form

$$\begin{aligned} K_H &= \delta H = 0 \\ K_S &= c^{-2}\delta p - \delta\rho = 0 \end{aligned} \tag{26}$$

(here $\delta = \partial_x + \lambda\partial_y$ represents variation along the characteristic line) which simply say that the enthalpy and entropy are constant along streamlines (even though they might vary from streamline to streamline).

On the other hand, $\lambda_\pm$ are only real for supersonic flow, in which case the associated characteristic lines $C_\pm$ are, respectively, left (upper) and right (lower) -running Mach lines inclined at an angle $\pm\mu = \pm\sin^{-1}(1/M)$ relative to the local streamline. The corresponding compatibility conditions are

$$K_\pm = \delta p \mp \frac{\rho}{\alpha}(v\delta u - u\delta v) \tag{27}$$

For irrotational flow, eq. (27) is equivalent to

$$\delta\theta = \pm\alpha\frac{\delta q}{q} \tag{28}$$

where $\theta$ is the angle that the local streamline makes with the $x$ axis. Eq. (28) can be integrated to give the following Riemann invariants:

$$R_{\pm} = \theta \mp \nu(M) \tag{29}$$

where $\nu(M)$ is the Prandtl-Meyer function [29].

For the adjoint equations, the eigenvalues and characteristic lines are the same as for the flow, and the associated compatibility conditions $r_{\lambda}^{T} A^{T} \delta\Psi$ can be obtained with the aid of an algebraic manipulation software (Wolfgram Research's Mathematica has been used in this work) and can be arranged in the following form

$$\begin{array}{ll} \lambda_0, & \begin{cases} \delta(\psi_1 - H\psi_4) = 0 \\ \delta\psi_1 + u\delta\psi_2 + v\delta\psi_3 + E_c\delta\psi_4 = 0 \end{cases} \\ \lambda_{\pm}, & \delta\psi_1 + u\delta\psi_2 + v\delta\psi_3 + H\delta\psi_4 = \pm\dfrac{1}{\alpha}(v\delta\psi_2 - u\delta\psi_3) \end{array} \tag{30}$$

(where $E_c = q^2/2$ ), which can be shown to agree with the results obtained in [26, 27].

For homentropic flows, the $\lambda_{\pm}$ compatibility conditions can be written as

$$\lambda_{\pm}, \qquad \delta(\psi_1 + u\psi_2 + v\psi_3 + H\psi_4) = \pm\frac{1}{\rho\alpha}\delta(\rho v\psi_2 - \rho u\psi_3) \tag{31}$$

The first compatibility condition in (30) can be immediately integrated, yielding the adjoint Riemann invariant $\psi_1 - H\psi_4$, which is known to be constant along streamlines (and everywhere for cost functions that only depend on the pressure [15]). Furthermore, in supersonic regions where the flow variables are constant, the above compatibility conditions give rise to the following 4 adjoint Riemann invariants $r_{\lambda}^{T} A^{T} \Psi$

$$\begin{aligned} R_1^{\psi} &= \psi_1 - H\psi_4 \\ R_2^{\psi} &= \psi_1 + u\psi_2 + v\psi_3 + E_c\psi_4 \end{aligned} \tag{32}$$

along streamlines,

$$R_{-}^{\psi} = \psi_1 + (u + v/\alpha)\psi_2 + (v - u/\alpha)\psi_3 + H\psi_4 \tag{33}$$

along right-running characteristics $C_-$ (with eigenvalue $\lambda_-$ ) and

$$R_{+}^{\psi} = \psi_1 + (u - v/\alpha)\psi_2 + (v + u/\alpha)\psi_3 + H\psi_4 \tag{34}$$

along left-running characteristics $C_+$ (with eigenvalue $\lambda_+$ ).

Even when they are not truly invariant, these Riemann functions have an additional interpretation as the adjoint variables corresponding to the flow compatibility conditions since

$$\Psi^T\nabla\cdot\vec{F}=\Psi^T A\partial_x U+\Psi^T B\partial_y U=\frac{2u}{q^2}R_2^\psi(\rho K_H-HK_S)+\frac{u}{q^2}R_1^\psi((H-E_c)K_S-\rho K_H)$$
$$+\frac{\alpha\left(u\alpha+v\right)}{2q^2}R_-^\psi K_- -\frac{\alpha\left(v-u\alpha\right)}{2q^2}R_+^\psi K_+ \qquad (35)$$

Hence, $R_{1,2}^\psi$ are related to perturbations along streamlines, while $R_\pm^\psi$ are related to perturbations along Mach lines. This insight will be used in the next section for the analysis of a particular example.

One last application of the Riemann functions concerns the jump conditions across characteristics. Across characteristic lines the adjoint solution can jump. When it does, the jumps $[\![\Psi]\!]_\lambda$ are subject to the conditions (20). The matrix $\hat{n}_\lambda\cdot(A,B)^T$ has rank 2 for streamlines and rank 3 for Mach lines. So for a streamline, there are 2 independent jump conditions, which correspond to the continuity of the Riemann functions $R_\pm^\psi$, while for a Mach line $C_+$ (resp. $C_-$) the 3 linearly independent jump conditions can be written in terms of the Riemann functions $R_1^\psi$ and $R_2^\psi$ and $R_-^\psi$ (resp. $R_+^\psi$), which is the Riemann function corresponding to the other Mach line. A visual summary of these results can be found in Figure 1.

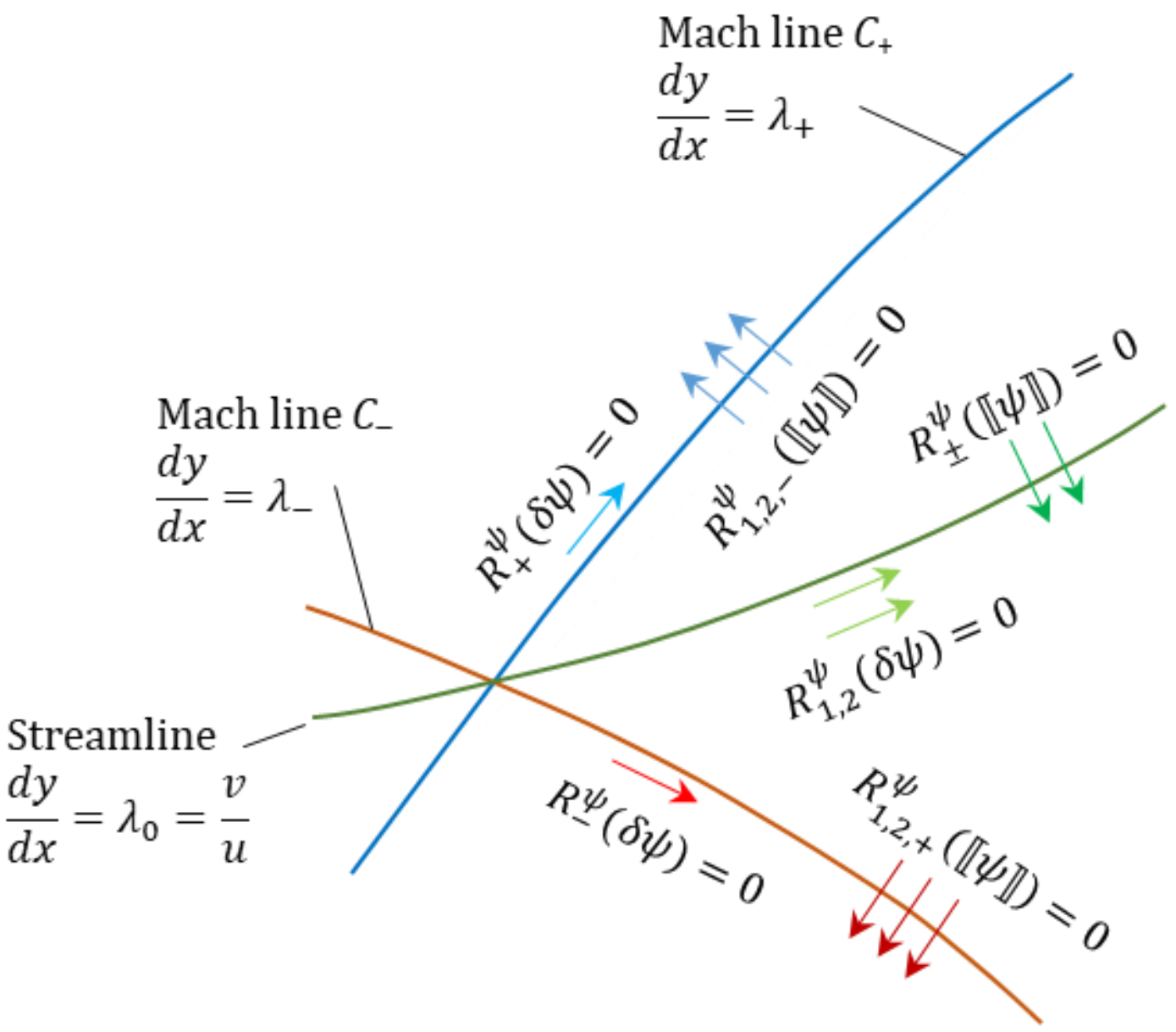

**Figure 1.**Sketch of characteristic lines and adjoint compatibility and jump conditions for supersonic flow. Each family of characteristics is represented with a different color. The number of arrows along and across the characteristic line represents the number of compatibility and jump conditions associated to that line.

### 2.5. *Extension to 3D*

In 3D, characteristic lines become characteristic surfaces (Mach conoids and streamline surfaces), defined by their normal vector $\hat{n}$ obeying [38]

$$\det(\hat{n} \cdot \vec{A}) = 0 \tag{36}$$

where $\vec{A} = \partial \vec{F} / \partial U$ are the flux Jacobian matrices. The corresponding flow and adjoint compatibility conditions can be written in the generic form

$$l_{\hat{n}}^{T} \vec{A} \cdot \nabla_{\hat{n}} u = 0 \tag{37}$$

and

$$r_{\hat{n}}^{T} \vec{A}^{T} \cdot \nabla_{\hat{n}} \Psi = 0 \tag{38}$$

where $l_{\hat{n}}$ and $r_{\hat{n}}$ are left and right eigenvectors obeying

$$l_{\hat{n}}^{T} (\hat{n} \cdot \vec{A}) = 0 = (\hat{n} \cdot \vec{A}) r_{\hat{n}} \tag{39}$$

and $\nabla_{\hat{n}} = \nabla - \hat{n}(\hat{n} \cdot \nabla)$ is the tangent gradient operator on the characteristic surface with normal vector $\hat{n}$. Eqs. (37) and (38) involve in general derivatives along two tangent directions, which make them cumbersome to use. In fact, conventional applications of the method of characteristics to 3D flows are numerical in nature and usually restricted to 2D or nearly 2D situations (axisymmetric or quasi-three-dimensional [39]) or conveniently restrict the analysis to two dimensions such as in the method of reference planes, where 2D characteristic lines are obtained from the projection of Mach cones and streamlines onto prescribed planes. However, on physical grounds it is known that at least two flow compatibility conditions (corresponding to entropy and enthalpy conservation) can actually be written as ODES along the streamline direction. In an analogous fashion, it was shown in [26] that the two 2D adjoint compatibility conditions that hold along streamlines (which in [26] where written as linear combinations of those shown in eq. (30)) also hold in 3D by simply adding the adjoint momentum variable along the third spatial direction. The remaining compatibility conditions will likely involve derivatives along two independent directions on the corresponding surface, the detailed analysis of which falls out of the intended scope of this paper and will not be explored further.

## 3. Application to a Supersonic Case: Flow Past a Diamond Airfoil

The method of characteristics can be used to solve for the flowfield in the case of steady, supersonic flow, and can be applied to the design of supersonic nozzles for 2D shock-free, isentropic flow [29], since in that case the non-linear flow equations reduce to algebraic equations along the characteristic lines. Since adjoint compatibility conditions cannot be integrated in general, there is little hope that a similar approach can be used with the adjoint equations. However, it turns out that the characteristic structure of the equations can be used to obtain analytic results in particular cases. One such example is given by the supersonic flow past a symmetric double wedge (diamond shaped) airfoil. The airfoil has a chord length of one, and a thickness of 0.06 (resulting in a half-angle of $\tau$ = 6.85 deg). The free-stream Mach number is $M_\infty$ = 2 and the angle of attack (AOA) is 0 degrees. The flow contains two sets of wedge-shaped oblique shocks attached to the leading and trailing edges, and two expansion fans emanating at the mid-chord vertices (Figure 2).

The flow solution can be computed exactly using shock-expansion theory [29]. The flow is parallel to the free stream upstream of the leading edge shock and downstream of the trailing edge "fishtail" shock. Between the leading edge shock and the leading Mach line of the expansion fan, the flow is parallel to the front segment of the airfoil, with a Mach number $M_1 < M_\infty$ that depends on the free stream Mach number and the wedge angle $\tau$, which also determine the shock inclination. The flow then turns across the expansion fan (with limit Mach numbers $M_1$ and $M_2$) so that behind the trailing Mach line of the fan the flow is parallel to the rear segment of the airfoil until it reaches the fishtail shock. The Mach number along the rear section $M_2 > M_\infty$ is determined by $M_1$ and the turning angle $2\tau$. Finally, the wedge angle, $M_2$ and $M_\infty$ determine the fishtail shock inclination. From shock-expansion theory, $M_1 \approx 1.755$ and $M_2 \approx 2.254$, which are in very good agreement with the numerical solution shown in Figure 3, computed with DLR's Tau solver [40] on an unstructured mesh with over 8.2 million nodes and 5300 nodes on the airfoil profile, with an outer freestream boundary domain of around 50 chord lengths from the geometry.

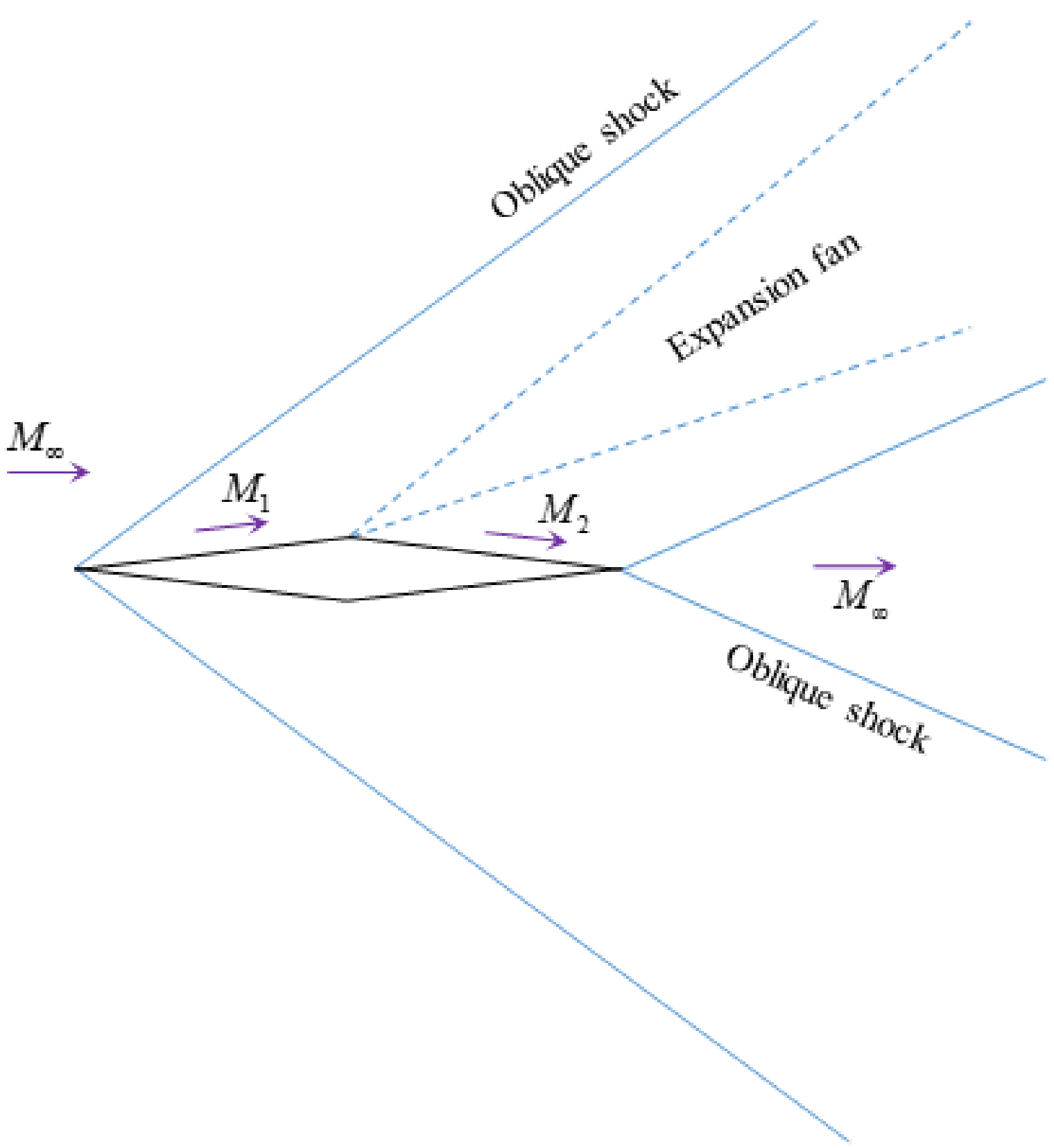

**Figure 2**. Sketch of the structure of supersonic flow over a diamond airfoil.

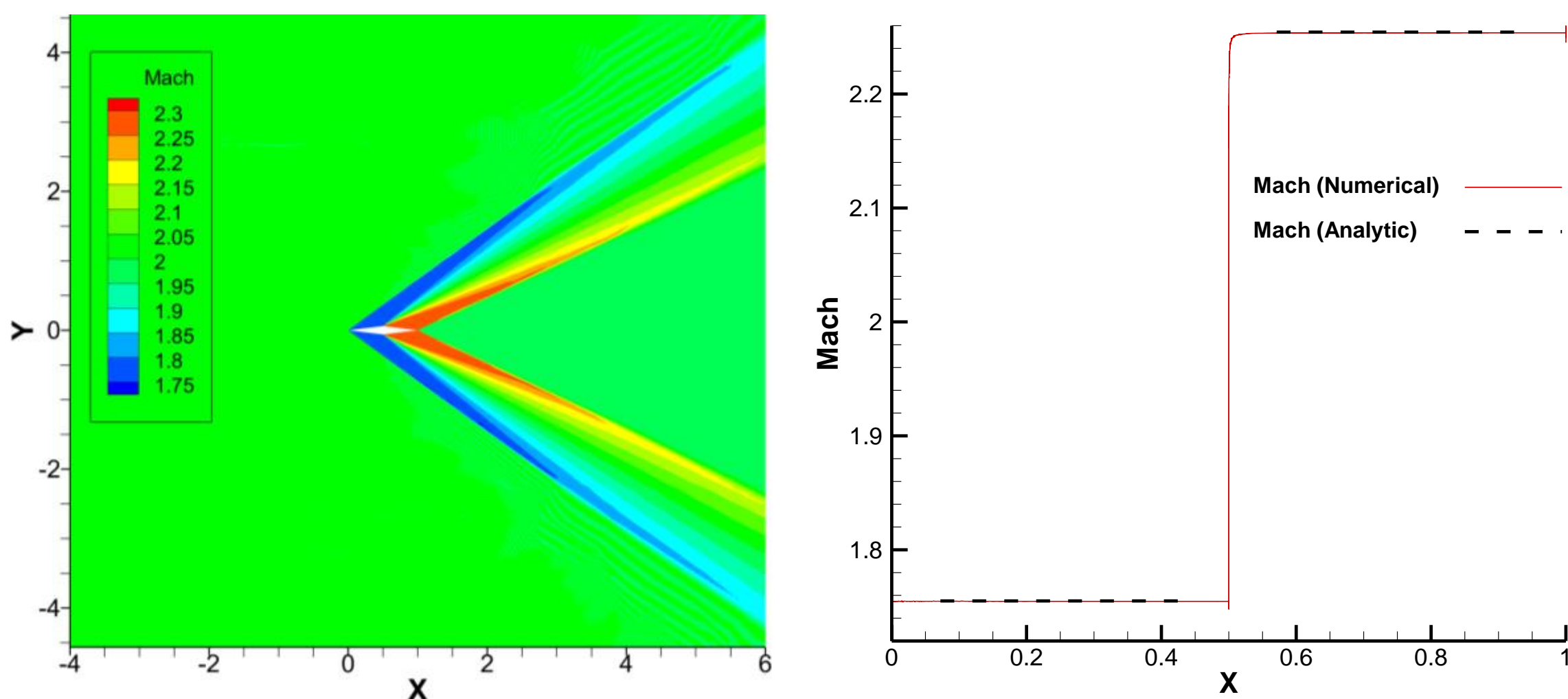

(**a**) (**b**)

**Figure 3**. Supersonic flow over a diamond airfoil with $M_\infty = 2$ and AOA = 0. Mach contours (**a**) and Mach along the airfoil profile (**b**) for a numerical solution computed with the Tau solver. Analytic Mach values obtained with shock-expansion theory are shown as black dashed lines.

The corresponding drag-based adjoint solution computed with Tau's discrete adjoint solver is shown in Figure 4. The adjoint variables are non-dimensionalized relative to Tau's reference values $(\rho_\infty, p_\infty, q_{ref} = \sqrt{p_\infty / \rho_\infty})$.

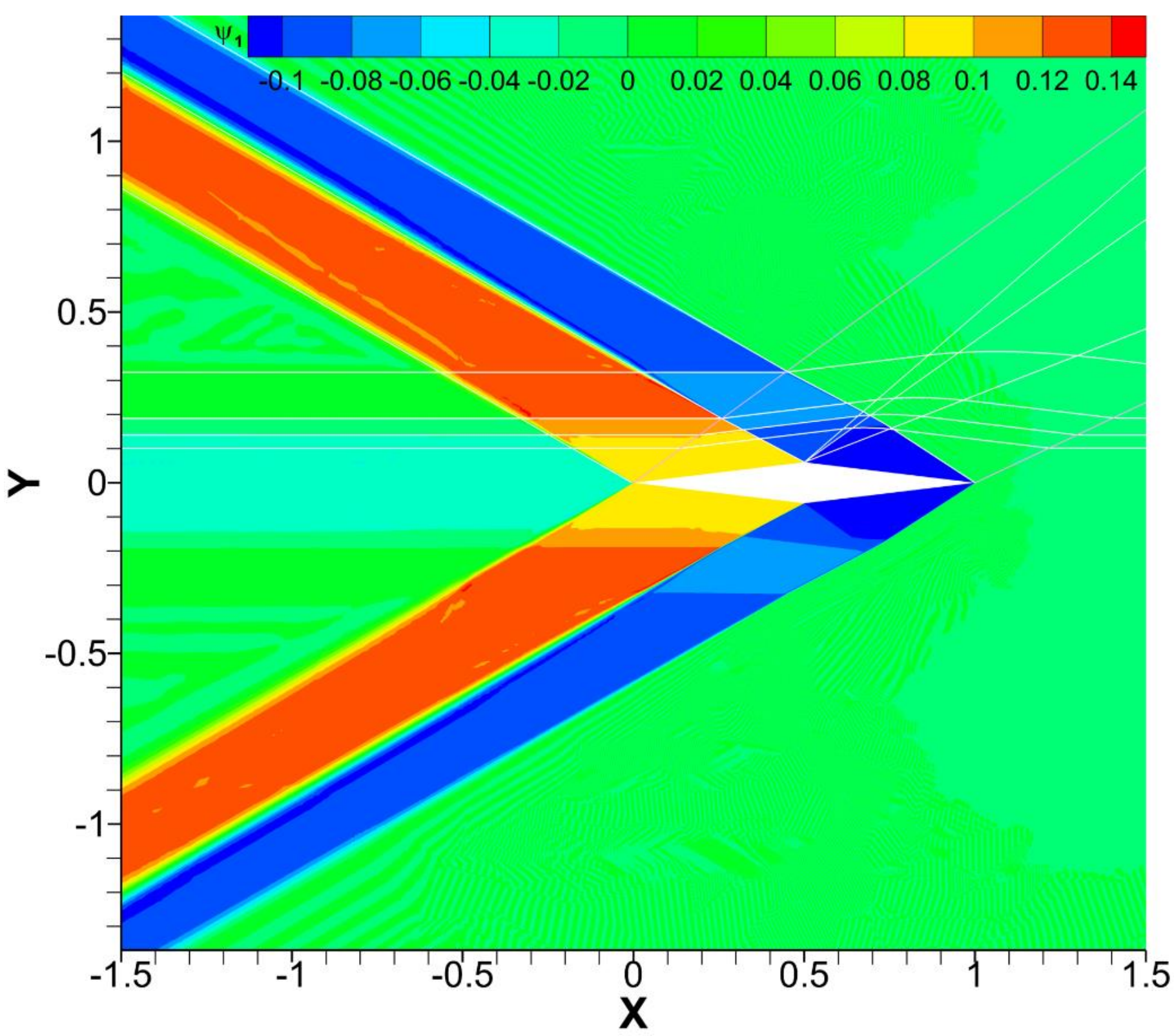

**Figure 4**. Supersonic flow over a diamond airfoil with $M_\infty = 2$ and AOA = 0. Contour plot of the drag adjoint density variable for a numerical solution computed with the Tau solver.

Figure 4 depicts the numerical adjoint solution along with the shocks and several notable characteristic lines (various streamlines, which are the nearly horizontal lines, as well as several Mach lines – the diagonal lines–, including the limits of the expansion fan and two Mach lines emanating from the trailing edge and running diagonally towards the upcoming flow). It can be seen that the adjoint solution clearly follows the characteristic structure of the flow in a pattern that somehow mirrors that of the primal flow. The adjoint solution vanishes downstream of the two Mach lines impinging on the trailing edge since no perturbation past those lines can affect the flow about the airfoil. The solution is essentially concentrated along strips limited by Mach lines emanating from the trailing edge, the mid-chord vertices and the leading edge, and the adjoint solution is discontinuous along these characteristic lines. There is also a weak horizontal strip along the incoming stagnation streamline upstream of the forward shock, in agreement with the general structure described in [41]. Finally, the solution along the wall is piece-wise constant (see Figure 5), and it is actually possible to use the jump conditions (20) and the wall boundary condition to predict the values of the adjoint variables.

In this particular case, the procedure starts at the trailing edge. On the upper side of the airfoil, the right-moving Mach line emanating from the trailing edge separates two regions where the flow and the adjoint solutions are constant. Hence, the Riemann functions are piecewise constant, and the only one that can jump is the one associated to that characteristic line. The other 3 are continuous across the line and, being constant on either side, maintain the zero value that they have downstream of the Mach line. Hence, the adjoint variables upstream of the Mach line obey the equations

$$\begin{aligned} &\psi_1 - H\psi_4 = 0 \\ &\psi_1 + u\psi_2 + v\psi_3 + E_c\psi_4 = 0 \\ &\psi_1 + (u - v/\alpha)\psi_2 + (v + u/\alpha)\psi_3 + H\psi_4 = 0 \end{aligned} \tag{40}$$

as well as the wall boundary condition eq. (8), which on the rearmost uppe r segment is simply $\psi_2 + \cot(\tau)\psi_3 = \kappa^{-1}$ . It is now easy to see that the above equations yield for the adjoint variables in the rearmost segment of the wall the following result

$$\begin{pmatrix} \psi_1 \\ \psi_2 \\ \psi_3 \\ \psi_4 \end{pmatrix} = \frac{\sin\tau}{\kappa\alpha_2} \begin{pmatrix} -\dfrac{(\gamma-1)M_2}{c_2} H_2 \\ \left(1+(\gamma-1)M_2^2\right)\cos\tau + \alpha_2\sin\tau \\ \alpha_2\cos\tau - \left(1+(\gamma-1)M_2^2\right)\sin\tau \\ -\dfrac{(\gamma-1)M_2}{c_2} \end{pmatrix} \tag{41}$$

where $\alpha_2 = \sqrt{M_2^2 - 1}$ and $c_2$ and $H_2$ are the speed of sound and total enthalpy at the rear part of the airfoil. Eq. (41) also holds on the lower side with the opposite sign for $\psi_3$. This analytic solution is compared with the numerical solution in Figure 5, showing excellent agreement.

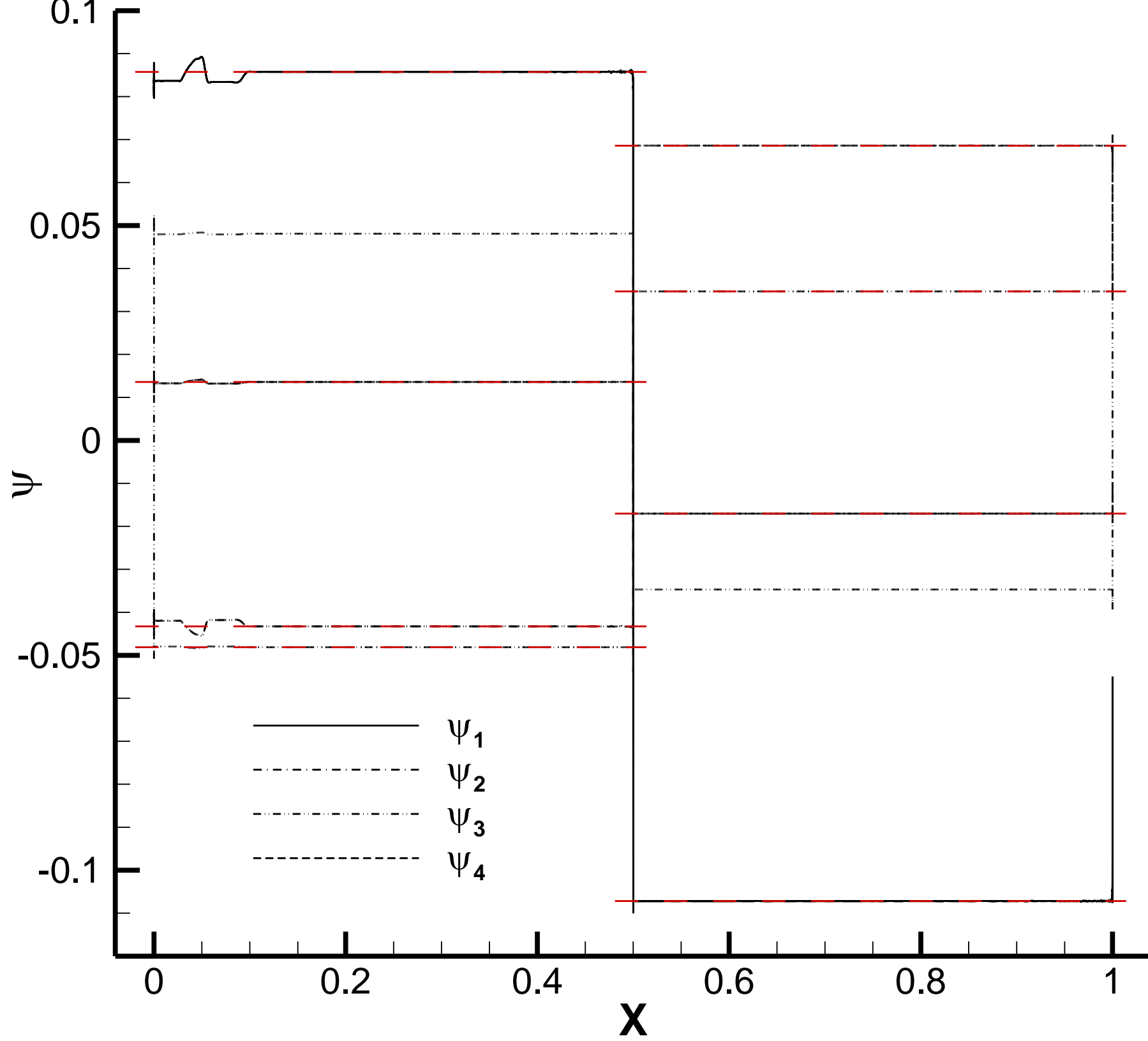

**Figure 5.** Supersonic flow over a diamond airfoil with $M_\infty = 2$ and AOA = 0. Adjoint variables along the wall for a numerical solution computed with the Tau solver. Analytic values as per eq. (41) and (59) are also shown as dashed red lines.

As can be seen in Figure 5, the adjoint solution along the wall is piecewise constant, with the values corresponding to the rear segment of the airfoil given by eq. (41). Away from the wall, the adjoint solution exhibits a fairly simple structure, at least in the immediate proximity of the wall, as can be seen in Figure 6, where the adjoint solution along a streamline is depicted and compared with the solution along the wall. The solution is again piecewise constant, and 5 plateaus (that will be labelled with roman numerals I-V) can be clearly identified, which are separated by jumps across characteristic lines. The adjoint variable is zero (V) beyond the right-running Mach line emanating from the trailing edge (7). Then it stays constant (IV) until the fan (III), delimited by Mach lines (3) and (5), where it has a continuous variation. It achieves a second plateau (II), which does not appear in the wall solution, and then jumps again across the right-running Mach line (2) emanating from the vertex of the fan [30, 33]. Finally, it jumps again across the right-running Mach line emanating from the leading edge (1). Notice that the plateau values I and IV agree fairly well with the wall values, which means that the adjoint solution is roughly constant throughout the corresponding regions.

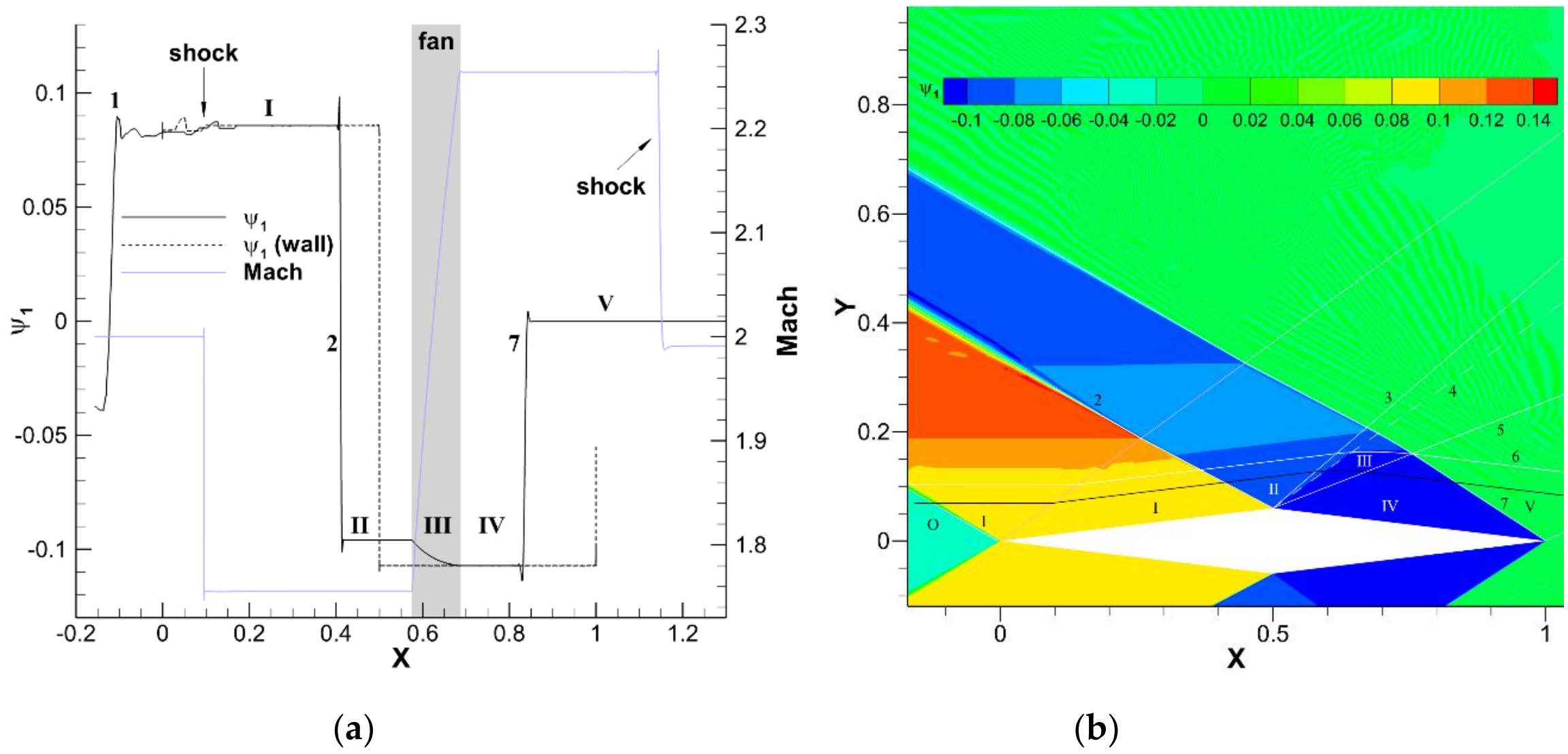

**Figure 6.** Supersonic flow over a diamond airfoil with $M_\infty = 2$ and AOA = 0. (**a**) First adjoint variable and Mach number along the (black) streamline indicated in panel (**b**) for a numerical solution computed with the Tau solver. The value of the adjoint variable along the airfoil profile is also shown for comparison. The plateaus are labelled I-V on the left panel and the corresponding regions are displayed on the right panel.

It is also interesting to note that the Riemann function $R_+^\psi$ (34), associated to left-running characteristics, is constant (and in fact vanishes) throughout the upper side of the airfoil. In fact, it is only non-zero in the region of the pressure side of the airfoil limited by the left-running characteristics emanating from the leading and trailing edges – see Figure 7. The reason for this behavior can be found in the interpretation of the adjoint Riemann functions as adjoint variables to the flow compatibility conditions. $R_+^\psi$ is the adjoint variable associated to left-running characteristics, and it is only non-zero in the region of the fluid domain where perturbations carried by left-running characteristics can reach the airfoil and, thus have an impact on the cost function.

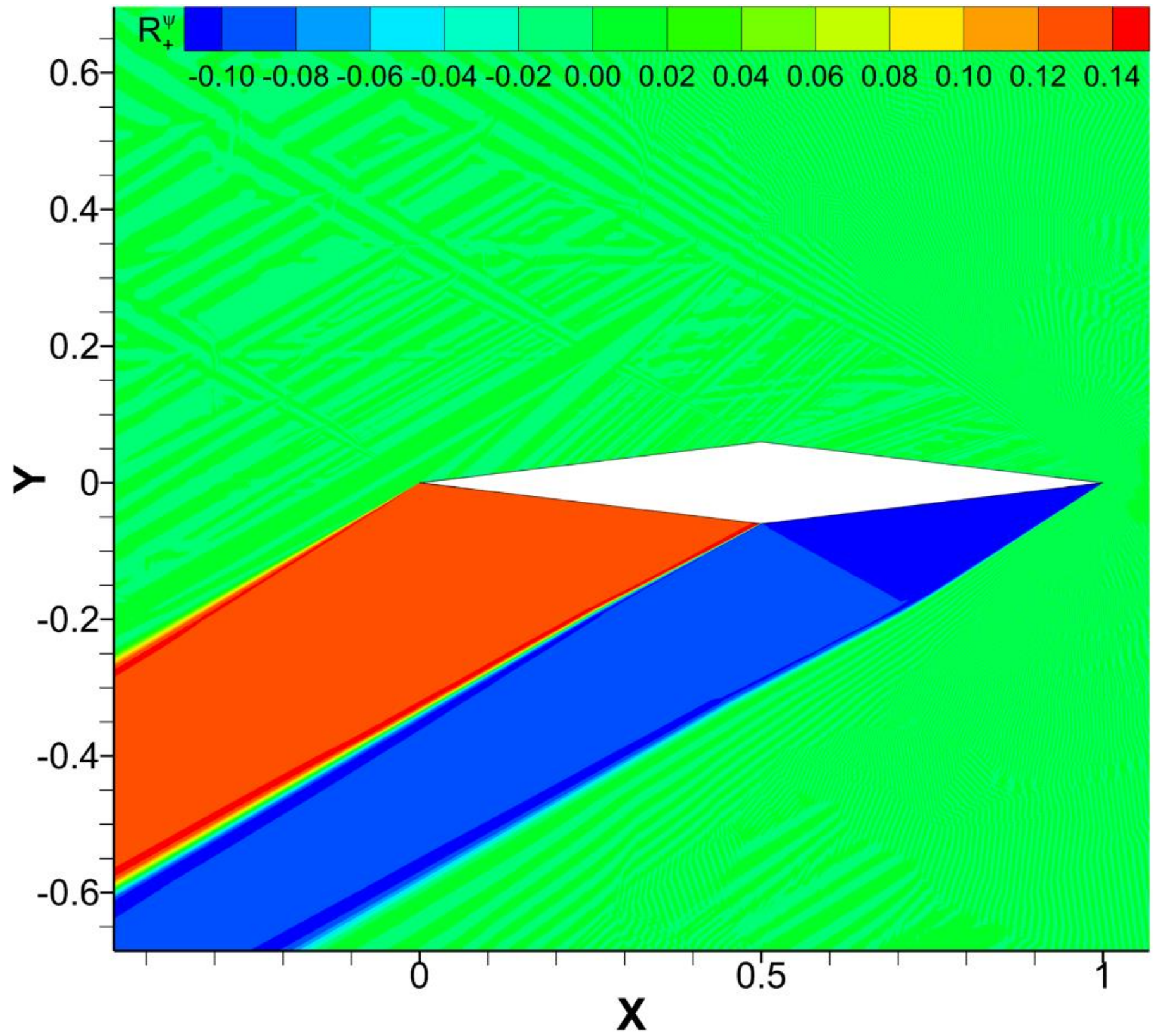

**Figure 7.** Contour plot of the Riemann function $R_+^\psi$ associated with left-running characteristics $C_+$ (with eigenvalue $\lambda_+$) for a numerical solution computed with the Tau solver. This function is roughly conserved along left-running characteristics (upward diagonal) and continuous across right-running (downward diagonal) characteristics.

This information can be used to extend the analytic solution to the forward part of the airfoil. Giles and Pierce's Green's function approach [17] allows to write the adjoint state (in fact, any 2D inviscid adjoint solution) in the following generic form:

$$\begin{pmatrix} \psi_1 \\ \psi_2 \\ \psi_3 \\ \psi_4 \end{pmatrix} = I_1 \begin{pmatrix} 1+\frac{\gamma-1}{2}M^2 \\ -\dfrac{u\left(1+(\gamma-1)M^2\right)}{q^2} \\ -\dfrac{v\left(1+(\gamma-1)M^2\right)}{q^2} \\ \dfrac{1+\frac{\gamma-1}{2}M^2}{H} \end{pmatrix} + I_2 \begin{pmatrix} 0 \\ -\dfrac{v}{q^2} \\ \dfrac{u}{q^2} \\ 0 \end{pmatrix} + I_3 \begin{pmatrix} -H \\ 0 \\ 0 \\ 1 \end{pmatrix} + I_4 \begin{pmatrix} -\dfrac{\gamma}{2}p_t M^2 \\ \gamma p_t M^2 \dfrac{u}{q^2} \\ \gamma p_t M^2 \dfrac{v}{q^2} \\ -\dfrac{\gamma}{2H}p_t M^2 \end{pmatrix} \tag{42}$$

where $p_t$ is the total pressure (which is constant between the leading and trailing shocks) and $I_j$, at any given point with coordinates $\vec{x}$, are the linearized cost functions corresponding to 4 linearly independent point source perturbations (mass, force normal to the flow direction, enthalpy and total pressure) [15] inserted at $\vec{x}$. For cost functions that only depend on pressure, $I_3 = 0$ (total enthalpy perturbation at constant pressure), while $R_+^\psi = 0$ requires that $I_2 = -\sqrt{M^2-1}I_1$. Using this in eq. (42) and rearranging, yields

$$\begin{pmatrix}\psi_1\\ \psi_2\\ \psi_3\\ \psi_4\end{pmatrix} = \frac{I_2}{Mc}\begin{pmatrix} -\dfrac{(\gamma-1)M}{c\alpha}H \\ \dfrac{1+(\gamma-1)M^2}{\alpha}\cos\theta-\sin\theta \\ \dfrac{1+(\gamma-1)M^2}{\alpha}\sin\theta+\cos\theta \\ -\dfrac{(\gamma-1)M}{c\alpha}\end{pmatrix} + I_4\begin{pmatrix} -\dfrac{\gamma}{2}p_t M^2 \\ \dfrac{\gamma p_t M}{c}\cos\theta \\ \dfrac{\gamma p_t M}{c}\sin\theta \\ -\dfrac{\gamma}{2H}p_t M^2\end{pmatrix} \tag{43}$$

This is the general analytic solution. It depends on two functions $I_2$ and $I_4$, whose values, for the numerical solution, are depicted in Figure 8.

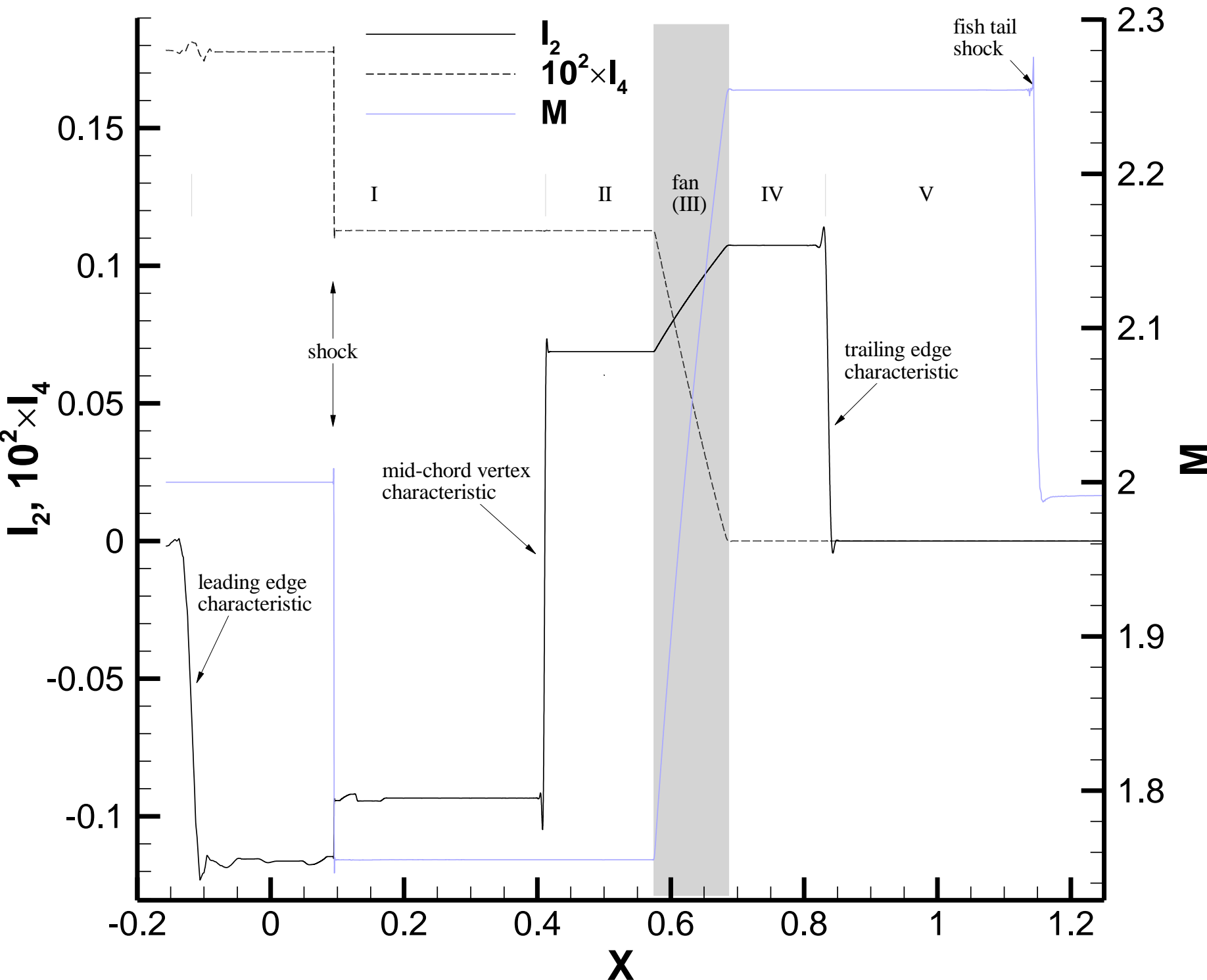

**Figure 8.** Supersonic flow over a diamond airfoil with $M_\infty$ = 2 and AOA = 0. $I_2$, $I_4$ and Mach number along the streamline indicated in Figure 6 for a numerical solution computed with the Tau solver.

In the extreme zones *I* and *IV*, $I_2$ is constant and its value is related to the adjoint wall boundary condition

$$I_2 = -v\psi_2 + u\psi_3 = Mc(\hat{n}_x\psi_2 + \hat{n}_y\psi_3) = Mc\hat{n}_x / \kappa \tag{44}$$

where $(\hat{n}_x, \hat{n}_y)$ is the unit wall normal vector pointing towards the fluid, yielding

$$I_2^I = -M_1 c_1 \sin\tau / \kappa \tag{45}$$

and

$$I_2^{IV} = M_2 c_2 \sin\tau / \kappa \tag{46}$$

respectively. Within the expansion fan (zone *III*) the flow is homentropic but variable, so along right-running characteristics eq. (31) applies. Using also $R_1^{\psi} = 0$, $R_+^{\psi} = 0$ and $R_{\pm}^{\psi} = I_1 \pm I_2 / \alpha$, where $I_1 = \psi_1 + u\psi_2 + v\psi_3 + H\psi_4$, yields

$$I_2^{III}(M) = \sigma \sqrt{\frac{\alpha(M)}{\rho(M)}} \tag{47}$$

where $\sigma$ is a constant. From eq. (47), the (constant) value in zone *II* can be written as

$$I_2^{II} = \sigma \sqrt{\frac{\alpha(M_1)}{\rho(M_1)}} \tag{48}$$

Lastly, in zone *IV*, $I_2^{IV} = \kappa^{-1} M_2 c_2 \sin\tau$, which fixes $\sigma$ as

$$\sigma = \frac{1}{\kappa} \sqrt{\frac{\rho_2}{\alpha_2}} M_2 c_2 \sin\tau \tag{49}$$

Focusing now on $I_4$, it is constant in zones *I*, *II* and *IV*, and actually $I_4^{IV} = 0$ since $[\![ I_4 ]\!] = 0$ across the right-running Mach line emanating from the trailing edge and $I_4 = 0$ downstream. Further information can be obtained by using the jump conditions across the right-running characteristic emanating from the midchord vertex to relate the solutions in zones *I* and *II*. In terms of the Riemann functions, the jump conditions are

$$\begin{cases} [\![\psi_1]\!] - H[\![\psi_4]\!] = 0 \\ [\![\psi_1]\!] + u[\![\psi_2]\!] + v[\![\psi_3]\!] + E_c[\![\psi_4]\!] = 0 \\ [\![\psi_1]\!] + (u - v/\alpha)[\![\psi_2]\!] + (v + u/\alpha)[\![\psi_3]\!] + H[\![\psi_4]\!] = 0 \end{cases} \tag{50}$$

The first and third equations in (50) are directly obeyed by the solutions on each side by construction, which leaves only the middle condition

$$[\![\psi_1]\!] + Mc\cos\theta[\![\psi_2]\!] + Mc\sin\theta[\![\psi_3]\!] + \tfrac{1}{2}M^2c^2[\![\psi_4]\!] = \frac{\gamma}{\gamma-1}\frac{M^2c^2}{2H} p_t [\![I_4]\!] = 0 \tag{51}$$

which yields $[\![ I_4 ]\!] = 0$ and, thus $I_4^I = I_4^{II}$, in agreement with Figure 8. Across the expansion fan, $I_4$ changes between $I_4^I = I_4^{II}$ and $I_4^{IV} = 0$. To obtain the value of $I_4$ across the fan, recall that $I_4$ is the effect on drag of a point perturbation to the stagnation pressure at constant static pressure [15]. Its value at a point $\vec{x}$ is given by the integration along the local streamline of secondary sources of mass and force normal to the local flow direction [18, 20]

$$I_4(\vec{x}) = -\frac{1}{\rho_t}\left(\int_0^{\infty} ds\partial_s q^{-2} I_1 + 2\int_0^{\infty} dsq^{-2}\partial_s\theta I_2\right) \tag{52}$$

In eq. (52), $\rho_t$ is the total density and the integral is taken from $\vec{x}$ ($s$ = 0) to the downstream farfield along the local streamline through $\vec{x}$. Let $\vec{x}$ be a point inside the expansion fan. Firstly, the contribution to the integral vanishes downstream of the trailing Mach line of the fan. Secondly, the flow is isentropic, so the following relations hold along any streamline:

$$\begin{cases} \partial_s \log q = \dfrac{1}{M\left(1+\frac{\gamma-1}{2}M^2\right)}\partial_s M \\ \partial_s\theta = \dfrac{1}{M\left(1+\frac{\gamma-1}{2}M^2\right)}\partial_n M \end{cases} \tag{53}$$

Furthermore, inside the fan the flow variables are constant along left-running Mach lines, so the following holds

$$\cos\mu\,\partial_s M + \sin\mu\,\partial_n M = 0 \Rightarrow \partial_n M = -\cot\mu\,\partial_s M = -\alpha\partial_s M \tag{54}$$

(here $\mu$ is the Mach angle). Gathering eq. (52) – (54) and using $I_1 = -I_2/\alpha$ yields

$$I_4^{III}(\eta) = -\frac{2}{\rho_t}\int_{\eta}^{\eta_2} d\eta \frac{I_2}{q^2 M\left(1+\frac{\gamma-1}{2}M^2\right)}(\alpha^{-1}-\alpha)\partial_\eta M \tag{55}$$

where $I_2$ is given by eq. (47) and the variable of integration has been changed to $\eta$, which is the angle that the local left-running Mach line makes with the $x$ axis and $\eta_2$ is the corresponding value for the final Mach line of the fan. The final piece of information required to evaluate eq. (55) comes from the analytic solution for the expansion fan [42]

$$M(\eta) = \sqrt{1+\frac{\gamma+1}{\gamma-1}\tan^2 z(\eta)} \tag{56}$$

where

$$z(\eta) = \sqrt{\frac{\gamma-1}{\gamma+1}}(\nu(M_1)+\tau-\eta+\pi/2) \tag{57}$$

Differentiating eq. (56) with respect to $\eta$ yields

$$\partial_\eta M = -\sqrt{\frac{\gamma+1}{\gamma-1}}\frac{\tan z}{M\cos^2 z} \tag{58}$$

Gathering all the information, the analytic adjoint solutions can be written as

$$\Psi_I = -\frac{\sin\tau}{\kappa}\begin{pmatrix} -\frac{(\gamma-1)M_1}{c_1\alpha_1}H_1 \\ \frac{1+(\gamma-1)M_1^2}{\alpha_1}\cos\tau-\sin\tau \\ \frac{1+(\gamma-1)M_1^2}{\alpha_1}\sin\tau+\cos\tau \\ -\frac{(\gamma-1)M_1}{c_1\alpha_1} \end{pmatrix} + \gamma p_t M_1 I_4^{III}(\eta_1)\begin{pmatrix} -\frac{M_1}{2} \\ \frac{\cos\tau}{c_1} \\ \frac{\sin\tau}{c_1} \\ -\frac{M_1}{2H_1} \end{pmatrix} \tag{59}$$

in zone $I$, where $\eta_1$ gives the inclination of the fan's leading Mach line relative to the $x$ axis,

$$\Psi_{II} = \frac{\sin\tau}{\kappa}\sqrt{\frac{\rho_2\alpha_1}{\rho_1\alpha_2}}\frac{M_2c_2}{M_1c_1}\begin{pmatrix} -\frac{(\gamma-1)M_1}{c_1\alpha_1}H_1 \\ \frac{1+(\gamma-1)M_1^2}{\alpha_1}\cos\tau-\sin\tau \\ \frac{1+(\gamma-1)M_1^2}{\alpha_1}\sin\tau+\cos\tau \\ -\frac{(\gamma-1)M_1}{c_1\alpha_1} \end{pmatrix} + \gamma p_t M_1 I_4^{III}(\eta_1)\begin{pmatrix} -\frac{M_1}{2} \\ \frac{\cos\tau}{c_1} \\ \frac{\sin\tau}{c_1} \\ -\frac{M_1}{2H_1} \end{pmatrix} \tag{60}$$

in zone $II$,

$$\Psi_{III} = \frac{\sin\tau}{\kappa}\sqrt{\frac{\rho_2}{\alpha_2}\frac{\alpha}{\rho}}\frac{M_2c_2}{Mc}\begin{pmatrix} -\frac{(\gamma-1)M}{c\alpha}H \\ \frac{1+(\gamma-1)M^2}{\alpha}\cos\theta-\sin\theta \\ \frac{1+(\gamma-1)M^2}{\alpha}\sin\theta+\cos\theta \\ -\frac{(\gamma-1)M}{c\alpha} \end{pmatrix} + I_4^{III}(\eta)\begin{pmatrix} -\frac{\gamma}{2}p_tM^2 \\ \frac{\gamma p_t M}{c}\cos\theta \\ \frac{\gamma p_t M}{c}\sin\theta \\ -\frac{\gamma}{2H}p_tM^2 \end{pmatrix} \tag{61}$$

across the fan and

$$\Psi_{IV} = \frac{\sin\tau}{\kappa}\begin{pmatrix} -\frac{(\gamma-1)M_2}{c_2\alpha_2}H_2 \\ \frac{1+(\gamma-1)M_2^2}{\alpha_2}\cos\tau+\sin\tau \\ -\frac{1+(\gamma-1)M_2^2}{\alpha_2}\sin\tau+\cos\tau \\ -\frac{(\gamma-1)M_2}{c_2\alpha_2} \end{pmatrix} \tag{62}$$

in zone $IV$, in agreement with eq. (41). The analytic solution eq. (59) – (62) is in remarkable agreement with the numerical solution as shown in Figure 5 and Figure 9.

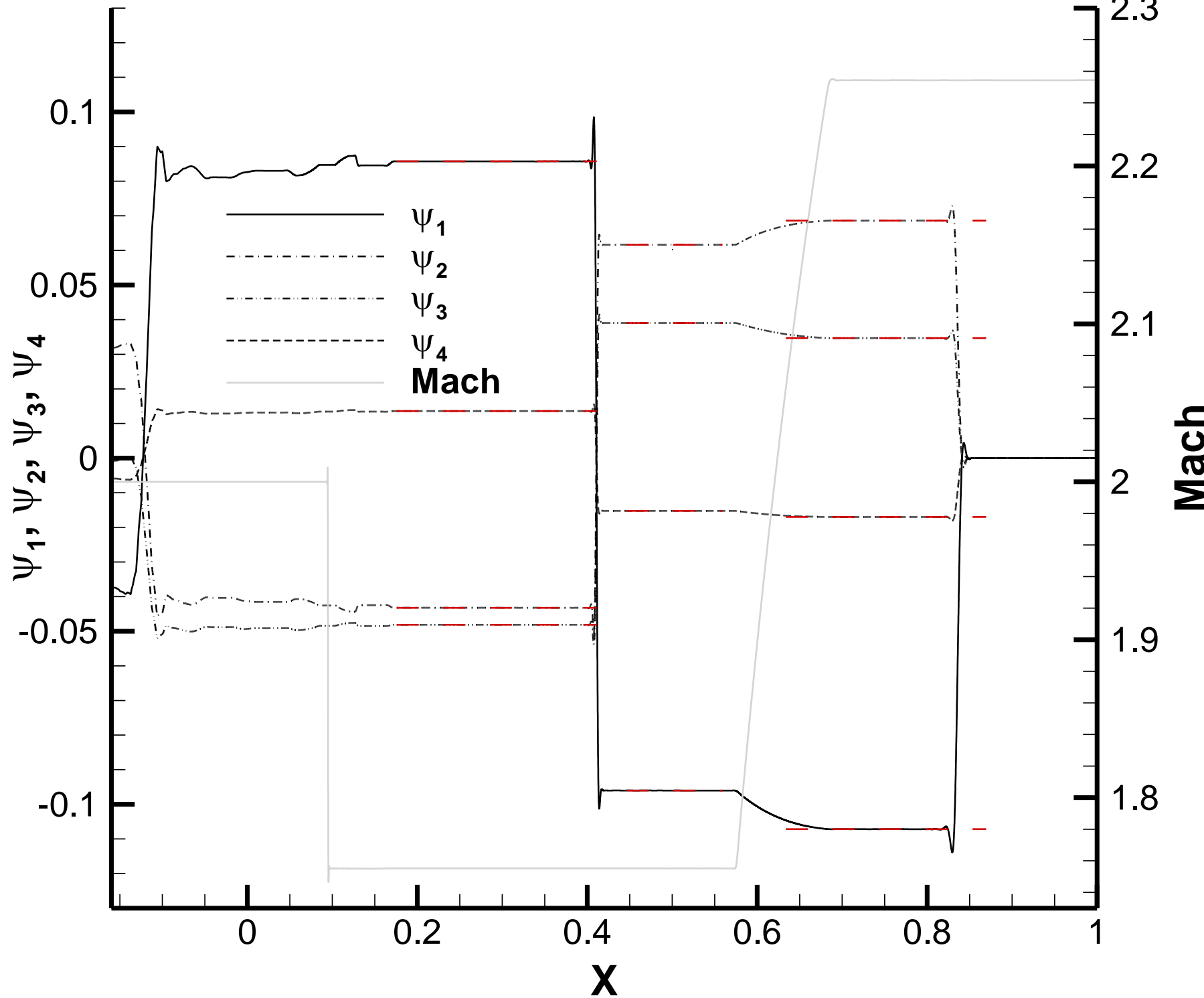

**Figure 9**. Supersonic flow over a diamond airfoil with $M_\infty = 2$ and AOA = 0. Adjoint variables and Mach number along the streamline indicated in Figure 6 for a numerical solution computed with the Tau solver. The analytic values for regions I, II and IV, as per eq. (41), (59) and (60), are shown as dashed red lines for comparison.

### 3.1. *Solution across the expansion fan*

Figure 10 sketches the fan region on the upper side of the airfoil. The figure shows the limiting left-running Mach lines of the fan (marked as 3 and 5), as well as the right-running Mach line emanating from the trailing edge (marked as 7), a streamline (6) impinging on the intersection of Mach lines 5 and 7 and an additional Mach line (4) picked at random within the fan.

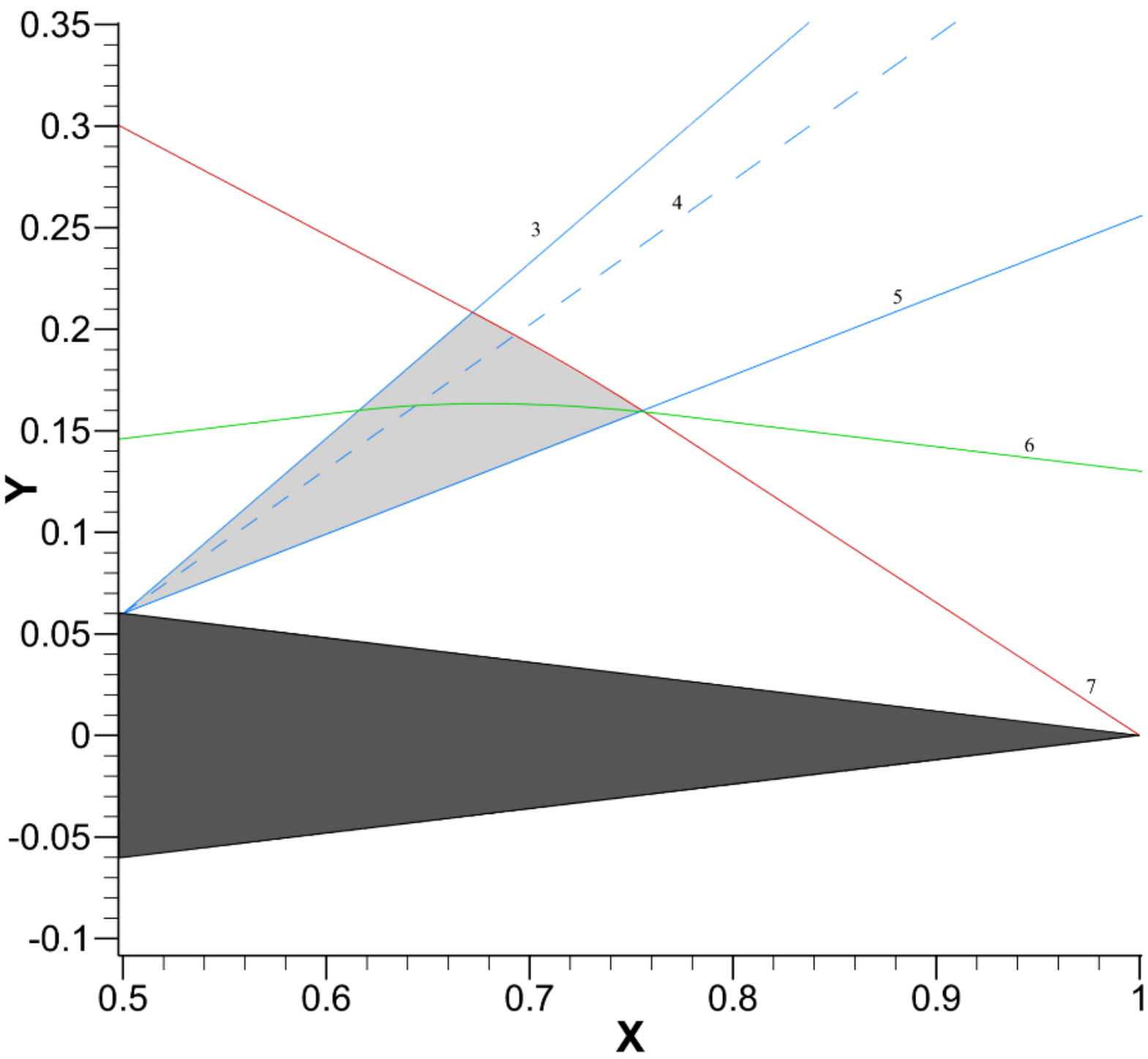

**Figure 10**. Supersonic flow over a diamond airfoil with $M_\infty = 2$ and AOA = 0. Sketch of the fan region. The rear part of the airfoil is shown in dark gray, while the light gray shaded region denotes the part of the fan where adjoint variables are different from zero.

Eq. (61) gives the spatially varying solution throughout the fan. Notice that, as is clear from eq. (61), the adjoint solution, like the flow solution, remains constant along each Mach line of the fan, as can be confirmed with the numerical solution shown in Figure 11, which depicts $I_2$ and $I_4$ and the adjoint variables along Mach lines 3, 4 and 5. The corresponding analytic values for $I_2$ and $I_4$ computed with eq. (47) and (55) are also shown for comparison, showing and excellent agreement.

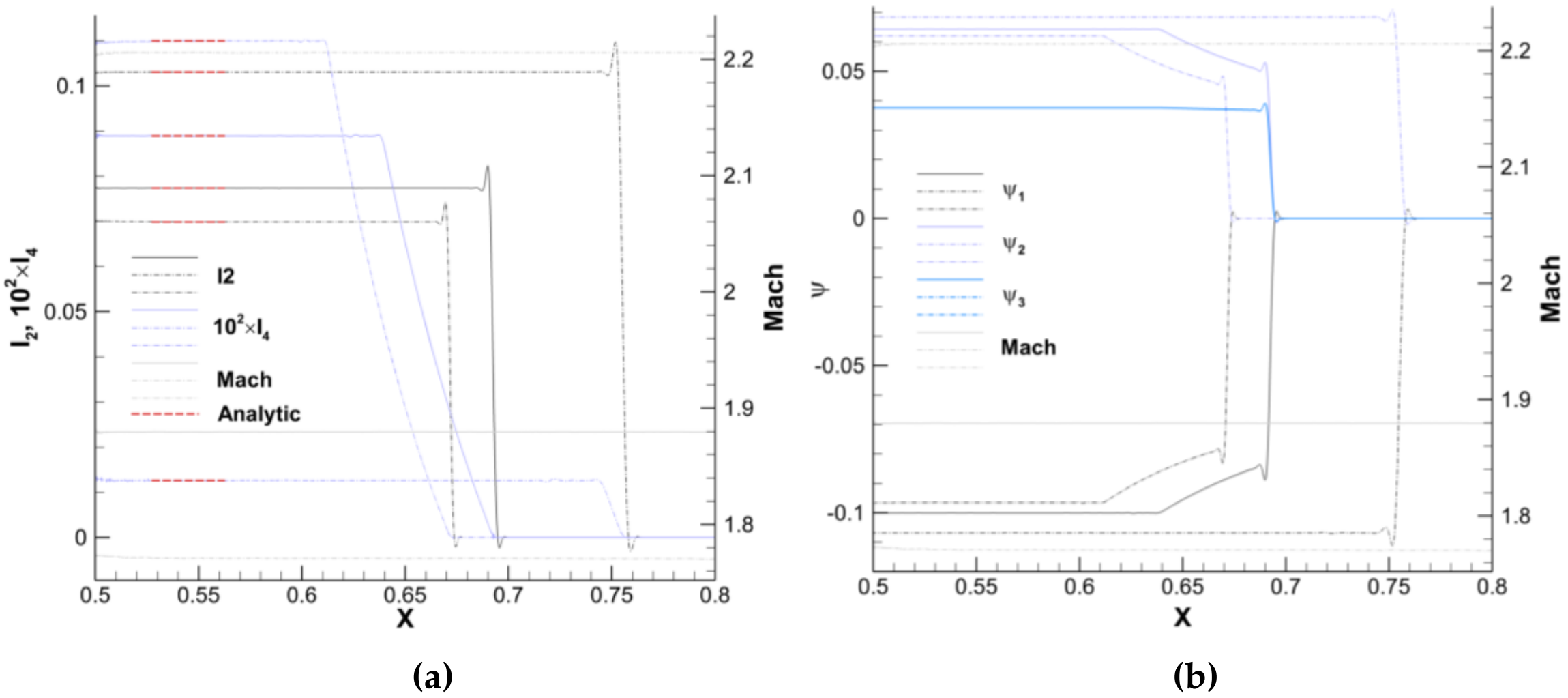

**Figure 11**. Supersonic flow over a diamond airfoil with $M_\infty = 2$ and AOA = 0. Plots along Mach lines 3 (dash-dot-dot), 4 (solid) and 5 (dash-dot) within the fan as indicated in Figure 6 and Figure 10 for a numerical solution computed with the Tau solver. (**a**) $I_2$, $I_4$ and Mach number. (**b**) adjoint variables and Mach number. Analytic values are depicted as dashed red lines.

It can be seen that towards the center of the fan (located at $x$ = 0.5 in the plot), $I_2$ and $I_4$ and the adjoint variables attain a constant value along each line of the fan, the value changing from line to line. $I_2$ and $I_4$ behave differently in this regard, since the former jumps abruptly across the right-running Mach line (7), while the latter shows a smooth variation and reaches a plateau at the intersection with the limiting streamline (6). The reason for this different behavior can be explained as follows. Both $I_2$ and $I_4$ can be written in terms of the adjoint Riemann functions as

$$
\begin{aligned}
I_2 &= \frac{\alpha}{2}(R_+^\psi - R_-^\psi) \\
I_4 &= \frac{1}{p_0}\left(\frac{\gamma-1}{\gamma} + \frac{2}{\gamma M^2}\right)R_2^\psi - \frac{1}{\gamma p_0 M^2}R_1^\psi
\end{aligned}
\tag{63}
$$

Both $R_+^\psi$ and $R_1^\psi$ vanish everywhere on the upper side of the airfoil, so it turns out that in the fan $I_2 \sim R_-^\psi$ and $I_4 \sim R_2^\psi$. Now, within the fan $R_-^\psi$ and $R_2^\psi$ are only different from zero in the shaded triangle bounded by Mach lines 3, 5 and 7. While $R_2^\psi$ (and, thus, $I_4$) is continuous across Mach line 7, $R_-^\psi$ (and, thus, $I_2$) is not, and its jump depends on the local value of the flow, which explains the difference between the plateau values along each line of the fan.

### 3.2. *Summary of the procedure*

The overall procedure can be summarized as follows. Focus on the upper side, since the solution on the lower side can be obtained by symmetry considerations. The adjoint solution vanishes downstream the characteristic line B emanating from the trailing edge (zone V in Figure 12) since no perturbation downstream of this line can affect the flow on the airfoil. The adjoint solution has four components that we choose to parameterize with four functions $I_1$, $I_2$, $I_3$, $I_4$ corresponding to the linearized drag due to four linearly independent point source terms injecting mass, transverse momentum, enthalpy and total pressure. Since the flowfield is piecewise constant (except across the fan), the functions $I_i$ are themselves piecewise constant, at least in the regions closest to the airfoil wall outside the fan. We divide the nearwall flowfield in 5 zones (labelled I to V in Figure 12). These zones are separated by the forward shock, two right-running Mach lines (labelled A and B in Figure 12) and the expansion fan delimited by left-running Mach lines C and D. A limiting streamline (labelled S and depicted in red) is also relevant since it sets a boundary for the adjoint solution within the fan. The solution procedure then amounts to finding the values of the $I_i$ on each zone using adjoint Riemann invariants and jump conditions. For pressure dependent cost functions, $I_3$ is identically zero everywhere. Besides, the adjoint left-running Riemann invariant $R_+^\psi = 0$ which yields $I_1 = -I_2/\alpha$, and $I_2$ is fixed by the adjoint wall boundary condition in zones I and IV. Finally, jump conditions across right-running waves entail $[[I_4]] = 0$, so $I_4$ is constant (and equal to zero) in zone IV and throughout zones I and II, where its non-zero value is fixed by the leading edge of the fan. The missing information is obtained by patching up the adjoint solution across the fan (zone III), where the solution is constant along each Mach line between the fan center and the limiting streamline S. The value of $I_2$ is determined using the compatibility conditions along right-running waves that cross the fan, while that for $I_4$ its definition as an integral along streamlines is used.

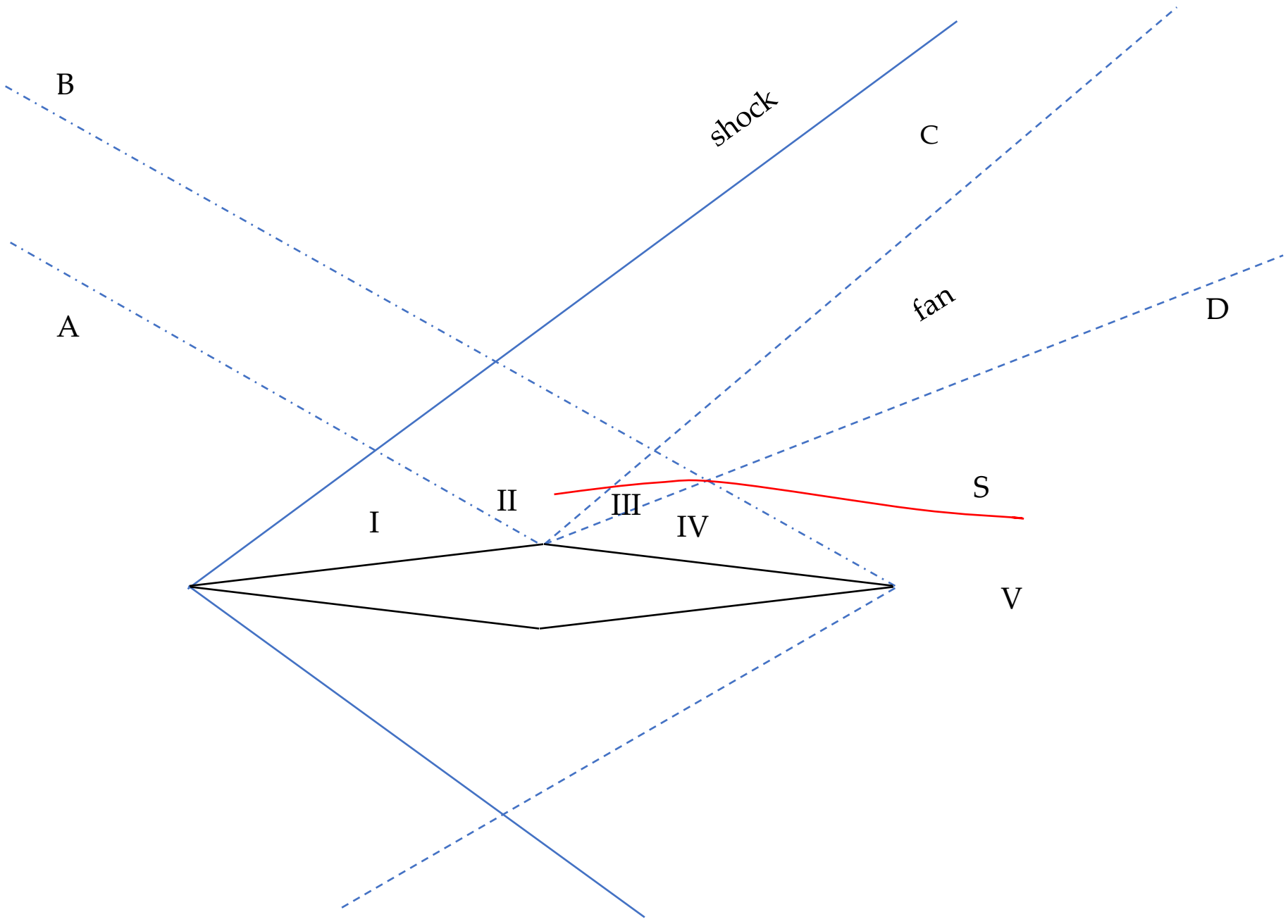

**Figure 12**. Sketch of the structure of the supersonic adjoint solution over a diamond airfoil. A, B, C, D and S are characteristic lines that separate different regions in the adjoint solution.

## 4. Conclusions

The behavior of the steady solutions to the 2D adjoint Euler equations is severely constrained by the characteristic structure of the equations, particularly in supersonic flows, where the equations are hyperbolic and the solutions display distinctive traits along certain significant characteristic lines.

In this work, the characteristic structure of the adjoint Euler equations in two dimensions has been examined. The eigenvalues and characteristic lines are the same as for the Euler equations, and compatibility conditions can be derived that constrain the evolution of the adjoint variables along the characteristics. At each point, these compatibility conditions are equivalent to the adjoint Euler equations. Additionally, adjoint solutions can be discontinuous along characteristic lines, and the jumps of the adjoint variables along these lines are constrained by jump conditions that can be used to build analytic adjoint solutions in simple cases, such as the supersonic flow past a diamond airfoil considered in this work.

The adjoint solution obtained in this paper has a relatively simple structure, at least in the vicinity of the airfoil profile. Due to the supersonic nature of the flow, the adjoint variables vanish downstream of two Mach lines emanating from the trailing edge. The near wall solution shows several constant patches delimited by the aforementioned trailing edge Mach lines, the expansion fan, a Mach line of the opposite family emanating from the center of the fan and two Mach lines emanating from the leading edge. The solution is parameterized in terms of the linearized objective functions corresponding to point sources of mass, transverse momentum, enthalpy and total pressure, whose behavior on the different zones and across their boundaries has been established using characteristic information. The obtained solution agrees extraordinary well with a numerical solution obtained on a very fine mesh.

The interest of the results described in this work is twofold. On the one hand, they provide benchmark solutions and constrains that can be used for verification of numerical adjoint solvers. On the other hand, by helping improve adjoint-based design tools, these insights may significantly impact methodologies used in the aerodynamic design processes and related applications in supersonic regimes, particularly as commercial supersonic flight becomes more relevant.

**Acknowledgments:** This research was funded by INTA under the grant IDATEC (IGB21001). The numerical computations reported in the paper have been carried out with the TAU code of the DLR (Deutsches Zentrum für Luft- und Raumfahrt), developed at the Institute of Aerodynamics and Flow Technology at Göttingen and Braunschweig. The Tau code has been licensed to INTA through a research and development cooperation agreement.

## Abbreviations

The following abbreviations are used in this manuscript:

| | |
|---|---|
| AOA | Angle of attack |
| ODE | Ordinary differential equation |
| PDE | Partial differential equation |
| 2D | Two dimensions/Two-dimensional |